# Observations of indirect exciton trapping in one- and two-dimensional magnetic lattices

A. Abdelrahman* and Byoung S. Ham**

*Center for Photon Information Processing School of Electrical Engineering, Inha University*
*253 Yonghyun-dong, Nam-gu, Incheon 402-751, South Korea*
* a.abdelrahman@inha.ac.kr ** bham@inha.ac.kr

(Dated: February 23, 2012)

**Abstract**

A simple method to create and control magnetic potentials onto coupled quantum wells is demonstrated for indirect-exciton magnetic confinement. Localized inhomogeneous magnetic potentials with periodically distributed local minima and maxima, known as magnetic lattices, are projected into the plane of the coupled quantum wells and used for the spatially distributed two-dimensional indirect-exciton trapping, in which case localized indirect-excitons are observed. The indirect-exciton trapping mechanism is examined by controlling external magnetic field bias resulting in a shift of the localized excitonic lattice.

In quantum heterostructures such as coupled quantum wells (CQWs), the presence of an artificially exerted confining potential quantizes the motion of quantum particles in discrete levels determined by the confining fields. Such confinement plays an important role for the control of quantum particles and opens a new direction to simulate condensed matter systems and to process quantum information in semiconductors. Interesting results have been recently reported for indirect-exciton trapping in CQWs using periodic electrical potentials [1-5]. In particular, trapping of indirect-excitons has been demonstrated by an electric potential for a single electrostatic trap [6,7], a one-dimensional electrostatic lattice [3], and two-dimensional electrostatic lattices [8,9]. Meanwhile, a special attention has been paid to magnetic field confinement as an alternative stable trapping mechanism for the indirect-excitons [10-13], where stable magnetic traps can be produced by using permanent magnetic materials [14]. Integrating magnetic materials with a quantum well system has been proposed theoretically [10,15,16] and demonstrated experimentally [13,17].

In this article we present a clear signature of stable magnetic confinement of indirect excitons in both one and two-dimensional magnetic lattices within a system of CQWs. The observed trapped indirect excitons are periodically distributed and controlled by shifting the trapping magnetic field local minima and maxima using modulating external magnetic bias field. The confinement stability of the inhomogeneous magnetic field is due to the use of permanently magnetized material to produce the trapping fields. The projected magnetic field local minima and maxima onto CQWs as shown in Fig. 1 are the origin of the magnetic confinement in a similar way to the magnetic trapping of ultracold atoms [14,18]. The magnetic field local minima $B_{min}$ are realized by fabricating specific patterns in a permanently magnetized thin film of thickness $\tau$, in which case the patterns extend through the thin film down to the surface of the underlying CQW layers. The presence of the patterns results in local field minima and maxima that appear at effective distances $d_{min}$ above the top and the bottom of the magnetized thin film, as shown in Fig. 1(c). The effective distance is defined as $d_{min} \sim \frac{\alpha}{\pi} \ln B_{int}$, where $\alpha$ represents both the length ($\alpha_h$) of each pattern and their separation ($\alpha_s$) between patterns, $B_{int} = B_0 \left(1 - \exp\left(-\frac{\pi}{\alpha}\tau\right)\right)$ with $B_0 = \frac{\mu_0 M_z}{\pi}$, and $M_z$ is the magnetization along the z-axis.

Thus, the $d_{min}$ is determined by the pattern parameter $\alpha$ ($\alpha_h$ and $\alpha_s$) and $\tau$. Depending on shape of the patterns the magnetic field local minima can be periodically distributed as simulated for the case of two-dimensional magnetic lattices in Fig. 1(d).

The center of each local field minima, i.e. the magnetic bottom of each single lattice site with $B=B_{min}$, is surrounded by relatively high fields known as magnetic barriers $\Delta B(k)$ which defines the space of the confinement and its depth $\Delta B(k) = |B_{max}(k)| - |B_{min}(k)|$. The curvature along the confining directions determines the trapping frequency $\omega_k$ which depends on the Zeeman levels of the trapped particles. For the case of harmonic potential, i.e. individual single site, the trapping frequency is given by $\omega_{k=x,y} = \frac{\beta}{2\pi}\sqrt{\mu_B g_F m_F \frac{\partial^2 B}{\partial k^2}}$, where $\omega_z = \sqrt{\omega_x^2 + \omega_y^2}$, $g_F$ is the Lande $g$-factor, $μ_B$ is the Bohr magneton, and $m_F$ is the magnetic quantum number of the hyperfine states [14]. Here we consider the ground state of the free indirect-excitons to be optically active resulting in recombination in a narrow radiative zone with a momentum close to zero [19]. As schematically represented in the Inset (1) of Fig. 1(b), the indirect-excitons become spatially trapped at their ground states in the bottom of each single trap (lattice site) represented as minimum magnetic potential perpendicular to the quantum-well growth direction (z-axis), Around each lattice site, where the magnetic field is relatively higher and almost parallel to the growth direction, the indirect-exciton's momentum increases at the cost of their energies. The difference in their energies defines the trapping energy. The exciton energy is defined as $B_{ext}(P) = -E_B \exp(-σ)I_0(-σ)$, where $I_0(-σ)$ is the modified Bessel function, $E_B$ is the binding energy, $σ = \left(\frac{Pl_B}{2\hbar}\right)^2$, and $l_B = \sqrt{\frac{\hbar c}{eB}}$ is the magnetic length [20]. For indirect excitons at nearly zero momentum, the magnetic length $l_B$ is comparable to the Bohr radius $a_B$ where in the case of short magnetic length $l_B < a_B$, the spin part of dominant indirect-exciton wavefunction is deformed and also causes the Zeeman splitting. For a small separation (<<$l_B$) between the electron and hole layers in CQWs subject to magnetic fields, the ground state of the system is resulted from the e-h interaction leading to a confined cloud of indirect-excitons that sufficiently extended in the confining space. In the present experiments, we fabricated the quantum well width less than $l_B$ (less than the Bohr radius of excitons in GaAs ~ 13.5 nm [21]). Thus, the indirect-excitons are regarded to mainly interact with the trapping inhomogeneous magnetic fields [20].

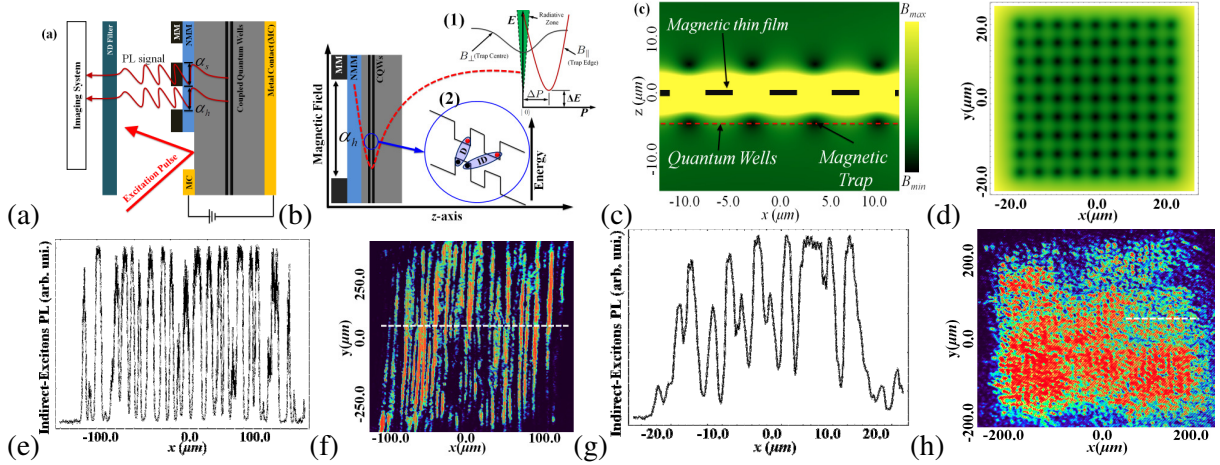

FIG. 1: (a) Schematic diagram of the experimental setup showing the integrated magnetic coupled quantum wells system. The excitation laser is set at ≈ 1000 μm away from the magnetic structure. (b) The existence of the hole in the permanent magnetic material creates a confining magnetic field with its minima projected into the location of the double quantum wells in which case the dispersion surface of the indirect excitons behaves according to the direction of the confining field as depicted in the inset (1). (c-d) The simulated magnetic field of a single two-dimensional magnetic lattice. (e-h) Experimental results of confined indirect-excitons in one and two-dimensional magnetic lattices, respectively. For 1D structure $α_h = α_s = 5$ μm with excitation pulse of $λ = 632$ nm and for 2D structure $α_h = α_s = 3.5$ μm with excitation $λ = 533$ nm.

The double quantum well system of Fig. 1(a) is grown by molecular beam epitaxy, where the sample consists of two 8 nm GaAs quantum wells separated by a 4 nm $Al_{0.33}Ga_{0.67}As$ barrier and surrounded by two 200 nm $Al_{0.33}Ga_{0.67}As$ for highly conducting layers. The metal contacts monitor the electric field along the z-direction for indirect-exciton generation. Using an rf sputtering technique a nonmagnetic material gadolinium gallium garnet (GGG) $Gd_3Ga_5O_{12}$ of thickness ≈3 μm is deposited on top of the CQW system, where the thickness of the nonmagnetic spacer is important for determining the effective distance $d_{min}$ and to allocate the magnetic field local minima and maxima within the quantum well layers. The permanent magnetic material $Bi_2Dy_1Fe_4Ga_1O_{12}$ is deposited with a thickness of ≈ 2 μm on top of the whole system (GGG + CQWs), as shown in Fig. 1(a).

We demonstrate magnetic trapping in two types of magnetic lattices (i) for one-dimensional magnetic lattices with dimensions $α_h=α_s=5$ μm as shown in Fig. 1(f) and with $α_h=10$ μm and $α_s=30$ as shown in Fig. 2, and (ii) two-dimensional magnetic lattices with $α_h=α_s=3.5$ μm as in Fig. 1(h). The $Bi_2Dy_1Fe_4Ga_1O_{12}$ is transparent to the light at wavelength $λ≥600$ nm [27], where the indirect exciton-generated photoluminescence is fully transmitted through the magnetic layer. The magnetic lattices for one and two-dimensional structures are fabricated by using focused ion beam lithography, annealed at 876.15 K, etched, and externally magnetized with magnetic field of 500 G before cooling to 13 K in a helium cryostat. In the case of the two-dimensional magnetic lattice, 3 x 3 blocks of 9x9 arrays of square holes are fabricated as shown in Fig. 2. For two different experiments, the sample is excited by HeNe laser at $\lambda$=633 nm and by Nd:YAG at $\lambda$=532 nm, respectively.

Indirect-excitons are known to have long lifetimes, orders of magnitude longer than direct-excitons, and can propagate longer in space at a bound $e$-$h$ state before recombined [3,28]. Here, the indirect-excitons are formed in and drifted to the vicinity of each single magnetic lattice via distinct photo-assisted capturing along with the injection of electrons and holes.

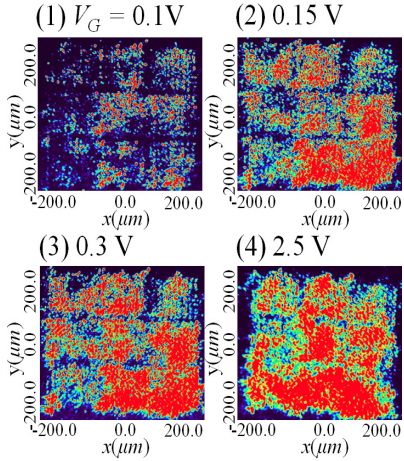

FIG. 2: Effect of increasing the injection rate of the electrons and holes via the potential gate voltage $V_G$, enhanced photoluminescence is observed at each single magnetic trap. The experiments are conducted for two-dimensional magnetic lattices with $\alpha_h = \alpha_s = 3.5$ µm using excitation pulse of HeNe laser with wavelength $\lambda = 632$ nm and power $P_{laser} = 1$ mW.

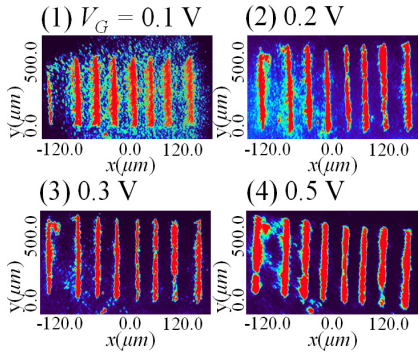

FIG. 3: The same enhanced photoluminescence of indirect-excitons is observed for one-dimensional magnetic lattice by increasing the injection rate of the carriers via the potential gate voltage $V_G$. The 1D magnetic lattice dimensions are $\alpha_h = 30$ µm and $\alpha_s = 10$ µm and HeNe laser excitation pulse with wavelength $\lambda = 632$ nm and power $P_{laser} = 1$ mW.

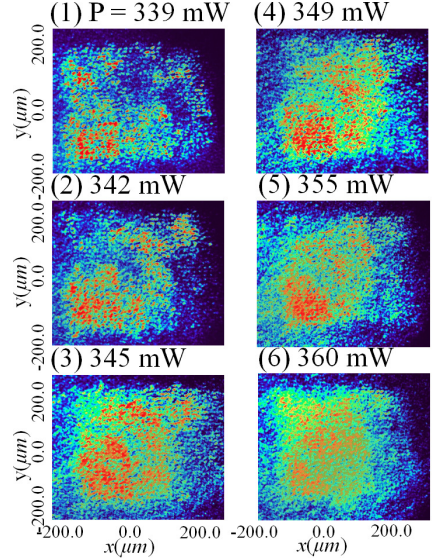

FIG. 4: Nd:YAG pump laser with wavelength $\lambda = 533$ nm is used to examine the effect of varying the excitation power $P_{laser}$ on the concentration of the trapped indirect-excitons for the case of two-dimensional magnetic lattice.

The gate voltage $V_G$, controls the electric field between the separated electron and hole layers. Increasing $V_G$ results in enhancement of the indirect exciton energy with respect to the e–h density [30]. Figure 2 shows enhanced photoluminescence as the gate voltage $V_G$ increases for both one and two-dimensional magnetic lattices.

As observed in Fig. 3, the indirect excitons are formed in the vicinity of each individual magnetic trap, where the density of the trapped clouds increases as the power of the excitation laser increases. It is merely because the environmental fluctuations and the indirect exciton scattering are smoothed out due to the magnetic confinement [31]. Fig. 3 shows enhanced concentration of the indirect-excitons as the laser power increases at $\lambda$=532 nm.

In this type of magnetic trapping, external magnetic bias fields are used to increase magnetic trapping field minimum value away from the zero minima to prevent the Majorana spin-flip process of the trapped particles [18]. However, the Majorana spin-flip process is avoided here because the zero point of the magnetic field local minima is allocated away from the formation plane of the indirect-excitons. The effect of the modulating external bias fields can be used as a clear signature to identify the magnetic trapping of the indirect-excitons as we discussed in Fig. 4 for two-dimensional magnetic lattices. We applied external magnetic bias fields, $B_{bias}\approx 75$ G, along the x-axis resulting in increased number of trapped indirect-excitons between the sites along the y-direction, as shown in Fig. 4(a). External magnetic bias field also causes a slight change in shape of the magnetic lattice, i.e. disappearance of the asymmetrical distribution of the sites [18], which is also observed in the electrostatic trapping of indirect-excitons [8,9]. It also causes a shift of the center of each individual site along the direction of the applied field. The site shift has been realized by spatially scanning across the centers of selected adjacent sites along the x-axis before and after applying the external bias field as presented in the Fig. 4(b).

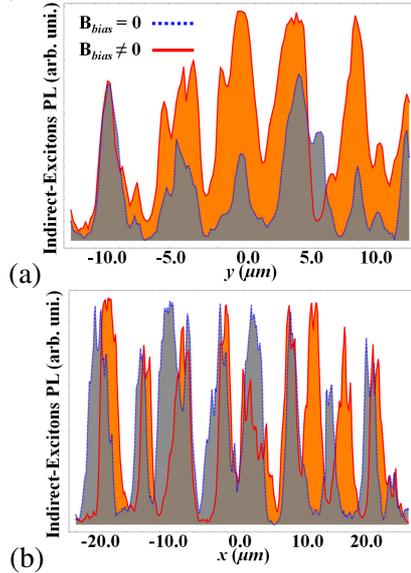

FIG.5: Applying external magnetic bias fields along the x-direction (a) increases the number of the trapped indirect-excitons along the y-axis and (b) shifts the center of each single trap (at the edges) along the x-axis.

To conclude, we observed magnetic trapping for indirect-excitons in one and two-dimensional magnetic lattices. Modulating external magnetic bias fields along the confinement direction may cause sufficiently cooled individual clouds of trapped indirect-excitons to tunnel between sites representing this approach as a suitable candidate for simulating condensed matter systems and processing quantum information.

**Acknowledgment**
This work was supported by the Creative Research Initiative Program (grant no. 2011-0000433) of the Korean Ministry of Education, Science and Technology via the National Research Foundation. We thank the Australian National University and Edith Cowan University in Western Australia for their help in fabricating the sample.